# Improving Josephson junction reproducibility for superconducting quantum circuits: junction area fluctuation

Pishchimova A.A.[1,2,a], Smirnov N.S.[1,a], Ezenkova D.A.[1], Krivko E.A.[1], Zikiy E.V.[1], Moskalev D.O.[1], Ivanov A.I.[1], Korshakov N.D.[1] and Rodionov I.A.[1,2,*]

[1] FMN Laboratory, Bauman Moscow State Technical University, Moscow, 105005, Russia
[2] Dukhov Automatics Research Institute, VNIIA, Moscow, 127030, Russia

Josephson superconducting qubits and parametric amplifiers are prominent examples of superconducting quantum circuits that have shown rapid progress in recent years. With the growing complexity of such devices, the requirements for reproducibility of their electrical properties across a chip have become stricter. Thus, the critical current $I_c$ variation of the Josephson junction, as the most important electrical parameter, needs to be minimized. Critical current, in turn, is related to normal-state resistance the Ambegaokar-Baratoff formula, which can be measured at room temperature. Here, we focus on the dominant source of Josephson junction critical current non-uniformity – junction area variation. We optimized Josephson junctions fabrication process and demonstrate resistance variation of 9.8–4.4% and 4.8–2.3% across 22×22 mm² and 5×10 mm² chip areas, respectively. For a wide range of junction areas from 0.008 μm² to 0.12 μm² we ensure a small linewidth standard deviation of 4 nm measured over 4500 junctions with linear dimensions from 80 to 680 nm. The developed process was tested on superconducting highly coherent transmon qubits (T1 > 100 μs) and a nonlinear asymmetric inductive element parametric amplifier.

## 1. Introduction

As superconducting quatum circuits complexity grows with anincrease of number of qubits, the tolerances for its frequency allocations become tighter in order to avoid frequency crowding [1] or to obtain the exact interaction frequencies, depending on the circuits architecture. Recent quantum processors and simulators already contain dozens and even hundreds of transmon qubits [2–9], which makes qubit frequency reproducibility one of the most important challenges. Transmon transition frequency follows $\hbar f_{01} \approx \sqrt{8E_c E_j} - E_c$, where $E_c = \frac{e^2}{2C}$ is charging energy and $E_j = \frac{\hbar I_c}{2e}$ is Josephson energy [10]. Qubit charging energy is accurately controlled, since capacitance $C$ depends on planar electrodes dimensions which can be reproducibly fabricated with modern microtechnology. Critical current $I_c$ at zero temperature is related to normal-state resistance $R_n$ by the Ambegaokar-Baratoff formula [11], which can be measured at room temperature. $I_c$ of the tunnel-junction, in turn, is defined by linear dimensions, tunnel barrier thickness and its microscopic structure. Typical nanoscale linewidth of Josephson junctions is challenging to reproduce with low variation using Dolan bridge technology. Any deviation of the linear dimensions of the tunnel barrier is directly reflected in the qubit frequency. Moreover, some qubit architectures, for example, fluxoniums or $0 - \pi$ qubits [12–14] require frequency tunability. In this case, parallel tunnel-junctions with significant asymmetry (50:1–70:1) are introduced in order to obtain the desired Hamiltonian parameters [15] and to reduce the sensitivity to flux-noise.

On the other hand, SNAIL parametric amplifiers [16] and SNAIL travelling wave parametric amplifiers (TWPA) [17] require long arrays of multiple high asymmetry non-linear elements, which determine their amplification range and saturation power. The described above high asymmetry features require reliable design and fabrication of Josephson junctions arrays with a wide range of areas (spanning from 0.01 to 1 μm²) on a single chip with the precise control of the small junctions area. One could increase the area of the junctions with decreased oxide thickness for better critical current control, but it leads to additional noisy two level defects [18].

In previous works, which described Al-AlO$_x$-Al junctions, suitable for superconducting qubit fabrication, 2.5–6.3% normal resistance variation was reported for 0.01–0.16 μm² junction area on 76-mm wafer [19]. 1.8% and 3.5% $R_n$ variation of 0.042 μm² junctions across 10×10 mm² chip and 49 cm² area respectively was demonstrated in [20]. Although impressive results were obtained, the authors did not control linear dimensions. The assessment of junction fabrication reproducibility was carried out solely by measuring the resistance of the SQUID without tight control of the smaller junction and therefore the asymmetry value was not controlled. It is worth noting, that fabrication of josephson junctions for superconducting qubit is rarely described in detail.

In this work we report on fabrication reproducibility improvement of small Josephson junctions (down to 80 nm), and investigate the origins of junction area non-uniformity. At first, we strived to reduce linear dimensions variance of small

a) Pishchimova A.A. and Smirnov N.S. contributed equally to this work

* Corresponding author: irodionov@bmstu.ru

junctions by optimizing e-beam lithography and get standard deviation of linewidth less than 4 nm for resist Dolan bridge features width from 80 to 680 nm. Secondly, we minimized junction resistance variation and edge roughness by optimizing the evaporation angle. As the result, we obtained 9.8–4.4% and 4.8–2.3% resistance variation for 22×22 mm$^2$ and 5×10 mm$^2$ chips respectively in a wide range of junction areas from 0.008 μm$^2$ up to 0.12 μm$^2$ using standard Niemer-Dolan technique. Furthermore, our research shows that there is a strong correlation of 0.82 between junction area and resistance, especially for small junctions of about 100 nm. We also show, that reproducibility is limited by evaporation system imperfection.

## 2. Materials and methods

The Josephson junctions (JJ), described in this work, were fabricated using Niemeyer-Dolan technique [21, 22]. This method does not require large evaporation angles, unlike Manhattan junctions, [23, 24] which lead to worse line edge roughness (LER). Another advantage is better suitability for fabrication of large arrays of junction, e.g., for fluxonium or parametric amplifiers, where junctions have to be placed directly one after another. However, this method has a poor Dolan bridge stability and narrow process window and, therefore, it requires precise control of technological operations. Recently developed substractive methods for junctions fabrication have shown promising results [25–28], but they require additional process steps with an extremely careful control of interfaces, which makes fabrication significantly more complicated. Nonetheless, the recommendations given in this paper can be used for all the three methods.

For this study, 1x1 inch$^2$ high high-resistivity silicon substrates ($> 10000\ \Omega \cdot cm$) were used. Next, a 100 nm thick epitaxial SCULL [29] Al base layer is deposited with ultra high vacuum (UHV) e-beam evaporation system. Probe pads were defined using a direct-laser lithography and then dry-etched in BCl$_3$/Cl$_2$ inductively coupled plasma. Resist bilayer for junction itself is composed of MMA(8.5)MMA copolymer (500 nm) and chemically amplified resist CSAR 62 (100 nm). Resist thicknesses were precisely controlled in order to minimize junction area variation. Resist thickness range $3\sigma$ is typically less than 5 nm. The josephson junctions are then defined using 50 kV electron-beam lithography tool. The development was performed manually in a bath of amilacetat at room temperature followed by a 60 second IPA dip and addittional 240 seconds in a IPA:DIW solution to get 200 nm undercut. We use oxygen descum followed by buffered oxide etchant (BOE) dip in order to remove e-beam resist residues and native oxide. Aluminum junction electrodes were e-beam shadow evaporated in a UHV system. The first electrode was evaporated at $\alpha_1$ angle, which we varied in our experiments and then statically oxidized at 5.2 mBar during 10 min. The second electrode was evaporated at $\alpha_2 = 0$ angle, see figure 1 for the cartoon of the process. Thicknesses of bottom and top electrodes are 25 nm and 45 nm, respectively. Resist lift-off was performed in a bath of N-methyl-2-pyrrolidone at 80°C for 3 hours and rinsed in a bath of IPA with sonication. Junctions normal resistance was measured with an automatic probe station.

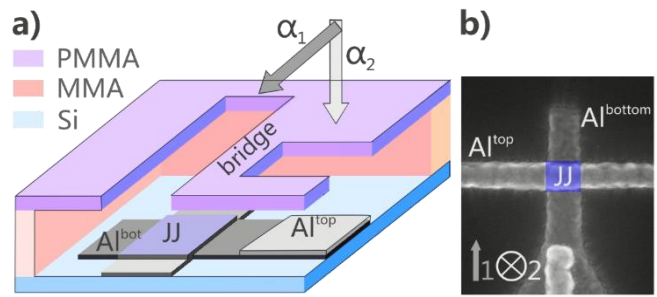

**Figure 1.** Niemer-Dolan bridge structure used in this paper. (a) Scheme of the used shadow evaporation process. Angular deposition was performed first, in order to avoid shading from the junction wall at the second deposition, that causes electrode discontinuity. Second junction was deposited at 0 angle. w$_1$ is top electrode mask mask linewidth. w$_0$ is Dolan bridge widh (b) SEM image of fabricated junction used in this work.

## 3. Experimental overview

The dominant source of junction resistance non-uniformity is junction area variations [20]. The deviation of the mask feature size could be minimized by optimizing e-beam lithography (EBL) process [30, 31]. There is typically a large junction wiring area (1–25 μm$^2$) leading to significant backscattering exposure of Dolan bridge during EBL. We simulated proximity effect using Monte Carlo method for the two different mask stacks: MMA-PMMA A4 and MMA-CSAR 62. The simulations indicate that the dose on the junction area increases by ~30% due to backscattering exposure, that corresponds to widening of the feature size by 50 nm in case of PMMA A4 top layer.The backscattering for high sensitive AR-P 6200.04 was 3 times lower. In the experiments we used high sensitive resist in order to minimize backscattering and fixed wiring design in order to keep linewidth (LW) biases constant.

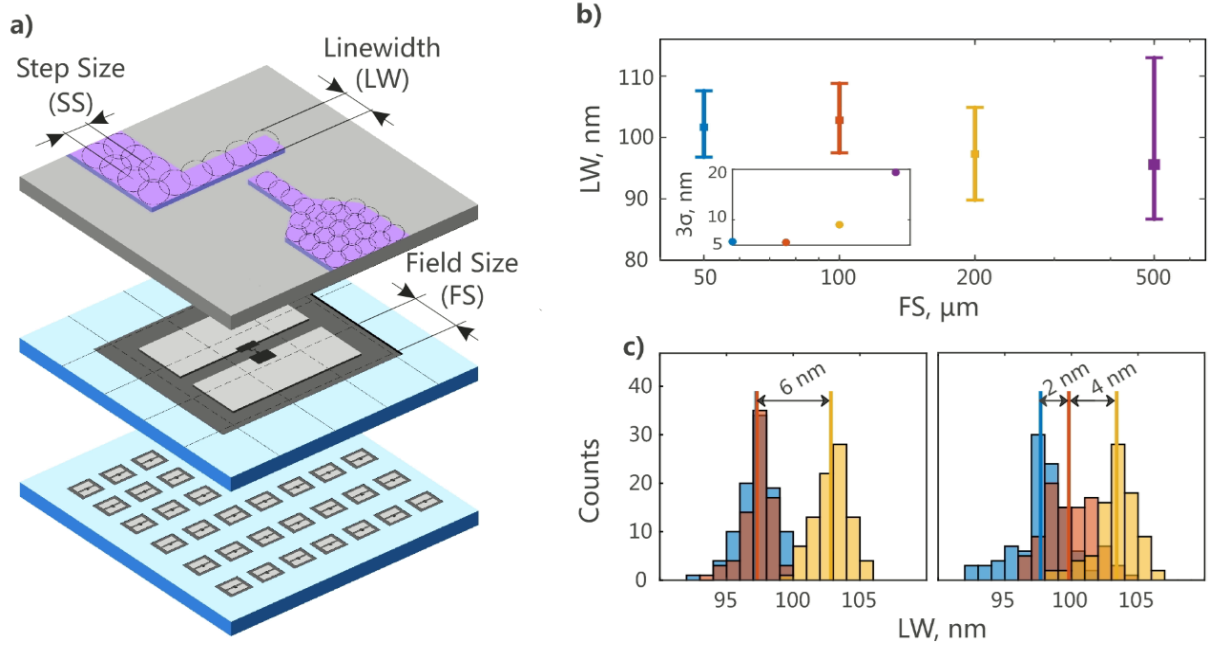

**Figure 2.** (a) Electron beam lithography process. From bottom to top: test chip, enlarged image of a single cell with a junction in the center, hidden image of the josephson junction in the resist. (b) Measured linewidth (LW) variation of the nominal 100 nm feature size grouped by field size. Inlet shows 3σ of linewidth variation. Dot color corresponds to the bars in the main picture. (c) Hystograms of measured linewidths for 100 nm (blue), 103 nm (red) and 105 nm nominal dimensions. When features are scanned along the junction (left picture), close nominal linewidth sizes are undistinguishable. Electrodes scanned across (right) have LWs difference corresponding to the beam step size.

First, we optimized e-beam lithography: writing field size and step size to study LW variations by measuring more than 1200 junctions. Direct SEM measurement of the resist mask LW is challenging, since the electron beam damages the resist mask and distorts the results. To overcome this problem, we instead measured the width of sputtered at 0° "shadows" of the mask features using an automatic SEM tool. Our test chip contained 98 junctions for each of the the patterning parameter: WF size (50, 100, 200, 500 μm), scanning direction (along and across), step size (2, 3, 5 nm) and linewidths: (100, 103, 105, 150, 300, 500 nm). The exposed area of the topology is divided into square write fields (WF) – maximum area that can be written at fixed stage position with the size. We minimized write field size (FS) in order to lower beam deflection and therefore lower aberrations, which increase LW variations and edge roughness. Cartoon of the layout exposure process is shown on figure 2a. With the smaller FS of 100 μm instead of 500 μm the linewidth 3σ variation decreases from 17.4 to 7.1 nm. For FS less than 100 μm the LW variation does not change (see figure 2b). We noticed the lowering of maximum positioning deviation from 41 nm to 33 nm for 500 μm and 200 μm field size, respectively, that may be explained by reduction of spherical aberrations.

Within the write field the pattern is scanned with a step size, which is limited by pattern generator frequency. The calculated step size equals to 2 nm for our 50 MHz tool with 180 μC/cm$^2$ and 200 pA beam current. We tried to obtain the highest accuracy corresponding to our step size. We did not observe any differences between exposed nominal feature sizes of 100 nm and 103 nm, when e-beam scanned the features along the junction electrodes. We used another scanning algorithm that scanned across the junctions and then the difference of 3 nm between feature sizes reproduced (figure 2c). The measurement results for the other nominal linewdths are listed in the supplementary materials.

Next, we experimentally investigated the influence of deposition angle on the mask LER. We automatically measured LER of the resist mask and the bottom electrode We used Matlab and ImageJ for image recognizion and LER quantification. Line edge roughness of the electrodes grows slowly from 2 nm for the deposition angles from 0° to 45° up to 4 nm, and dramatically increases starting from 45° (figure 3a). Such a big LER values can affect the $I_c$ variation of comparably small junctions (sub 100 nm), which are common for many devices. After the first junction electrode deposition at an angle, there is a rough metal layer deposited on the resist mask walls (figure 3b), which increases LER of the resist mask. This increased resist mask roughness transfers into the second deposited electrode. However, smaller evaporation angle could lead to unstable overlay of the electrodeswhich results in additional resistance variation (figure 3d). We experimantally compared two different evaporation angles $\alpha_1$ of 40° with full overlay (figure 3e) and 35° with partial overlay (figure 3f) by measuring 5000 Josephson junctions with areas from 0.008 to 0.12 μm$^2$. There is the factor of 2 improvement in resistance spread for the electrodes with full overlay. Out theory is that there are slightly different evaporation angles across the substrate due to evaporation system imperfectness. The margin for junction electrodes overlap compensate this angle variation. The overlay margin can be increased by reducing the width of the bridge, but minimal bridge width is limited by its stability. We used 150 nm wide bridge in our experiments, as it provides robustness for the whole range of junction areas.

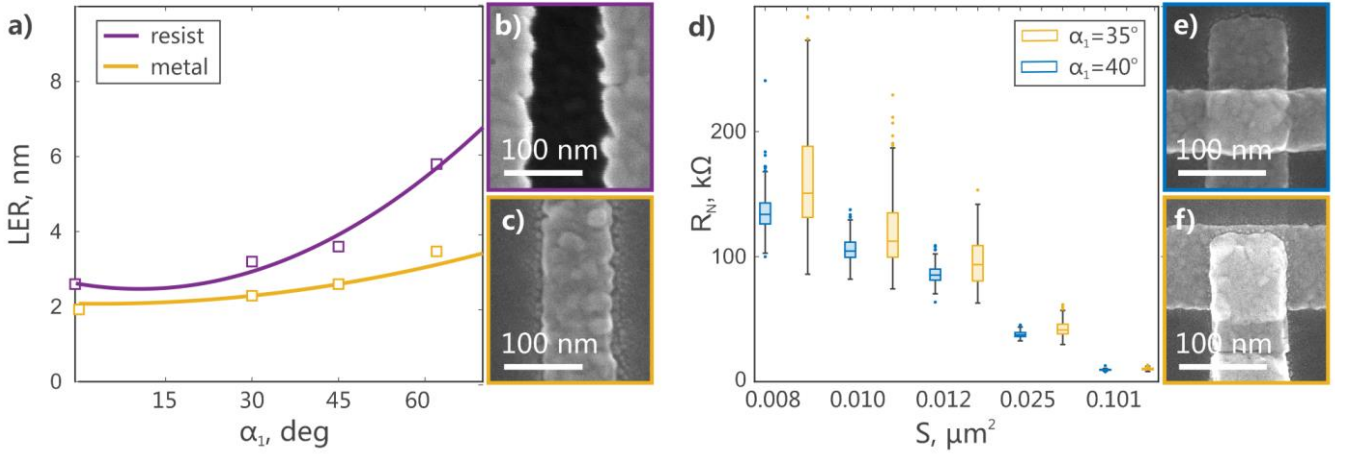

**Figure 3.** (a) Line edge roughness of resist mask after evaporation (purple) and bottom electrode (yellow) as a function of deposition angle (b) SEM image of resist mask after evaporation at $\alpha_1 = 62°$. (c) SEM image of measured bottom electrode evaporated at at $\alpha_1 = 62°$. (d) Measured normal resistances grouped by nominal junction area. Blue color corresponds to junctions evaporated at 40° angle with electrodes fully overlayed. Yellow – junctions evaporated at 35° with incomplete electrodes overlay. (e) and (f) show SEM images of 0.01 μm² junctions used in the experiment with and without full overlay respectively.

## 4. Results

In order to quantitively investigate JJ reproducibility across the substrate, we fabricated 22×22 mm² chip with statistically significant number of Josephson junctions (> 2500) and measured their normal resistance at room temperature. The substrate contained six 5×10 mm² chips with differenct junctions areas: 0.008, 0.010, 0.012, 0.025 and 0.120 μm². Figure 4a shows heat maps of measured resistances and LWs across the sample, fabricated with the following parameters: 100 nm field size, 2 nm step size, 45° evaporation angle. Resistance spread across the substrate is 4.4–9.8% for JJ areas 0.120–0.008 μm², the average inter-chip variation is 3.1–6.3%. Furthermore, the average variation is 2.3–4.8% across single 5×10 mm² chip. Substrate-scale chip LW standart deviation σ for 150×170 nm² junctions is about 4 nm. (the heat maps of the other junction areas can be found in supplementary materials) Figure 4b shows normal resistance for the directly measured JJ areas. The slope of the linear fit is -1, since $R = 1/A$., where A is the area of the juncion. The correlation coefficient between normal resistance and JJ area is 0.82, thus the main contributor in $R_n$ fluctuation is LW variation. We speculate that the other source of resistance spread is oxidation inhomogeneity. We observe a linearly dependent gradient of resistance on the heat map, which is also reproduced on LW heat map. We assumed that the gradient of $R_n$ originates from the evaporation source imperfectness. To test this we have fabricated another substrate, with the both electrodes evaporated at 0° angle to eliminate any significant angle variations. LW variation gradient became less visible. The LW standart deviation σ at 0° angle was lower than at 45°: 3.3 nm against 4 nm (the heat map is presented in the supplementary materials). To summarize, our Junctions resistance variation is mostly limited by junction area fluctuations, which, in turn, are limited by evaporation system imperfectness and resolution of our 50 keV e-beam lithography tool.

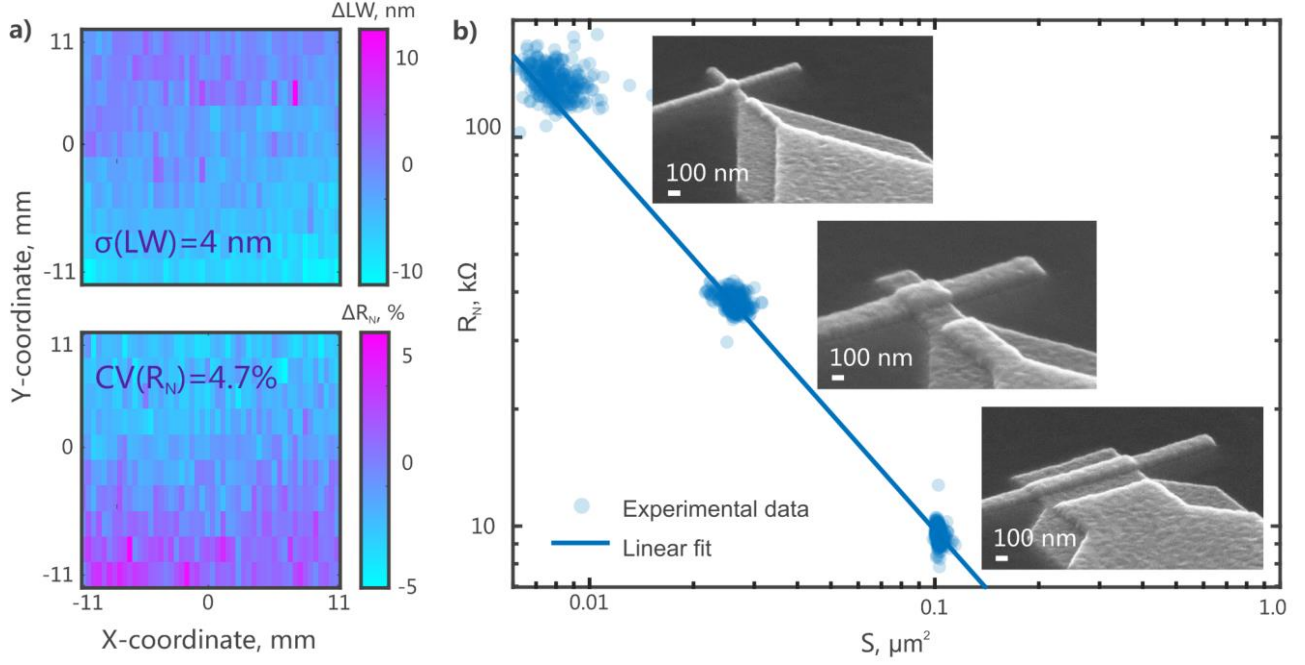

**Figure 4.** (a) Heat maps of measured top electrode LW and resistance $R_N$ for 150x170 nm$^2$ junctions. Both on the LW map and the resistance map there is a visible gradient originated by evaporation source imperfection. (b) Measured resistance vs. measured junction area. Junctions were fabricated in three area groups: 0.008, 0.025 and 0,12 μm$^2$. Solid line is a linear fit. SEM images of the junction in the corresponding junction area groups. SEM-images show fabricated Josephson junctions with corresponding 0.008, 0.025 and 0,12 μm$^2$ areas.

To test the developed process, we fabricated and measured two similar chips with 6 single transmon qubits each, using the optimized lithography and evaporation parameters for junctions, described in this work. We defined capacitances and microwave lines the same way as the probe pads in this work. We also deposited bandages to shortcut the parasitic junctions [32] and placed free-standing airbridges across microwave lines in order to supress the parasitic modes. Two qubits on the chip were frequency-fixed, the other had various SQUID asymmetries α (0.25, 0.35, 0.45, 0.55). The areas of 250x260 nm$^2$ for single junction and 200x220 nm$^2$ for the larger junction of the asymmetrical SQUIDs were used. The chips were wire-bonded in the Cu boxes and measured in dilution refrigerator at 20 mK. Coherence and frequency masurements are presented in Table 1. Next, we fabricated and tested SNAIL parametric amplifier with 25 SNAIL cells and large asymmetry. The areas of 0.63 μm$^2$ and 0.055 μm$^2$ for large and small junctions respectively were used (α = 0.087) in order to obtain 6 GHz frequency.

Table 1. Fabricated devices properties

| Device | Parameter | Average |
| --- | --- | --- |
| Qubit chip 1 | T1, μs | 129.3±19.5 |
|  | T2*, μs | 45.7±18.6 |
|  | $f_{01}$, GHz | 4.46±0.12 |
|  | $\Delta f_{01}$, % | 2.06 |
|  | Δα, % | 8.0 |
| Qubit chip 2 | T1, μs | 190.4±67.5 |
|  | T2*, μs | 16.7±10.0 |
|  | $f_{01}$, GHz | 4.32±0.08 |
|  | $\Delta f_{01}$, % | 1.37 |
|  | Δα, % | 10.5 |
| Parametric amplifier | Δf, % | 3.2 |
|  | Δα, % | 3.3 |

## 4. Conclusions

In an effort to improve Josephson junction reproducibility, we have performed a systematic study of technological origins of critical current non-uniformity, focusing on junctions area fluctuations. To do this, we fabricated a signifiant amount of junctions and directly measured their dimensions and normal resistance. High sensitive resist stack was picked in order to lower the backscattering exposure. We minimized e-beam lithographer write field to improve junctions linewidth variation and used proper scanning algorithm to increase accuracy. Evaporation angle showed a direct effect on junctions line edge roughness. LER of the resist mask transfers to the deposited junction and grows with increasing evaporation angle. We provided full junction electrodes overlay and improved resistance variation in comparison to incomplete overlay. The developed fabrication process demonstrated 9.8–4.4% and 4.8–2.3% resistance variation for 22×22 mm2 and 5×10 mm2 chips respectively in a wide range of junction areas from 0.008 μm$^2$ up to 0.12 μm$^2$. We found a strong corellation of 0.82 between normal resistivity and junctions area. Junction reproducibility is dominantly limited by resist mask linewidth fluctuations and could be further improved by using lithography tool with higher resolution. We have also noticed that the evaporation system imperfectness distorts junction area and therefore increases $I_c$ variation.

## Acknowledgements

Technology was developed and samples were made at the BMSTU Nanofabrication Facility (FMN Laboratory, FMNS REC, ID 74300).

# Supplementary material for
# «Improving Josephson junction reproducibility for superconducting quantum circuits: junction area fluctuation»

**Resistance and linewidth measurements**

Test junctions were defined into resist bilayer of 500 nm MMA(8.5)MAA copolymer and 100 nm of CSAR 62 using followed exposure parameters: 100 um field size and 2 nm step size (see Materials and methods). Next, we deposited 50 nm of Al at 0° angle, performed lift-off and measuref "shadows" of the mask features using an automatic SEM tool. Our test chip contained 98 junctions of different nominal top electrode linewidths: 100, 103, 105, 150, 300, 500 nm. The measurement results for different scanning algorithms (across the electrode and along the electrode) are presened in Table 1.

*Table 1. Measurements of resist mask linewidth*

| Scanning direction | along | | | | | | across | | | | | |
|---|---|---|---|---|---|---|---|---|---|---|---|---|
| $LW_{nom}$, nm | 100 | 103 | 105 | 150 | 300 | 500 | 100 | 103 | 105 | 150 | 300 | 500 |
| LW, nm | 99 | 99 | 104 | 150 | 302 | 502 | 96 | 99 | 101 | 144 | 300 | 500 |
| 3σ, nm | 4,4 | 4,4 | 5,1 | 5,5 | 6,0 | 6,6 | 4,5 | 4,4 | 5,8 | 8,3 | 5,6 | 5,2 |
| N | 98 | 98 | 96 | 97 | 95 | 97 | 93 | 94 | 84 | 88 | 89 | 98 |

The heat maps of both top and bottom electrodes linewidths and junctions resistance across the 22x22 mm$^2$ area are presented in Table 2. Mean measured value with 99.7% confidence limit is shown on the corresponding heat map. We obtained the gradient in resistance, which is also reproduced in top electrode linewidth. We assumed that the gradient originates from evaporation source imperfectness. To test this we have fabricated another substrate and evaporated both electrodes at 0° angle to eliminate any significant angle variations. The heat maps of features linewidths are shown in Table 3. The gradient in the top electrode linewidth was no longer visible and the variation was lower in comparison to the electrodes evaporated at an angle.

*Table 2. Measurements of fabricated josephson junctions*

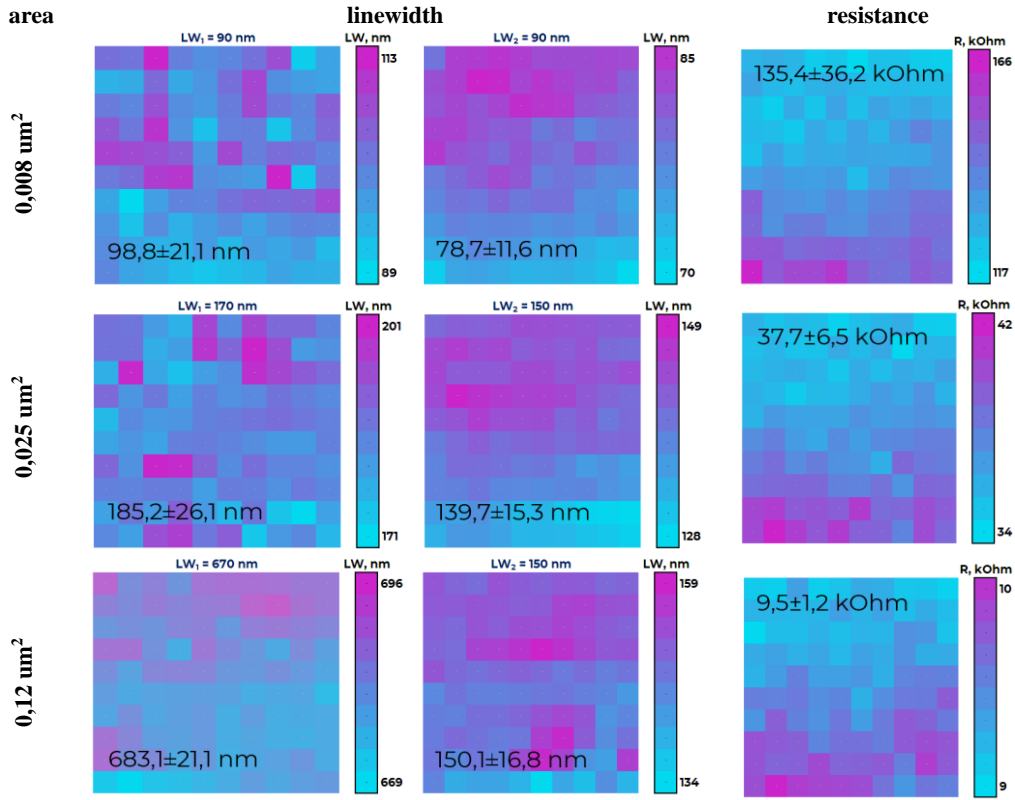

*Table 3. Measurements of resist mask linewidth*

| area | Heat map |
|---|---|
| **0,008 um²** | |
| **0,025 um²** | |
| **0,12 um²** | |

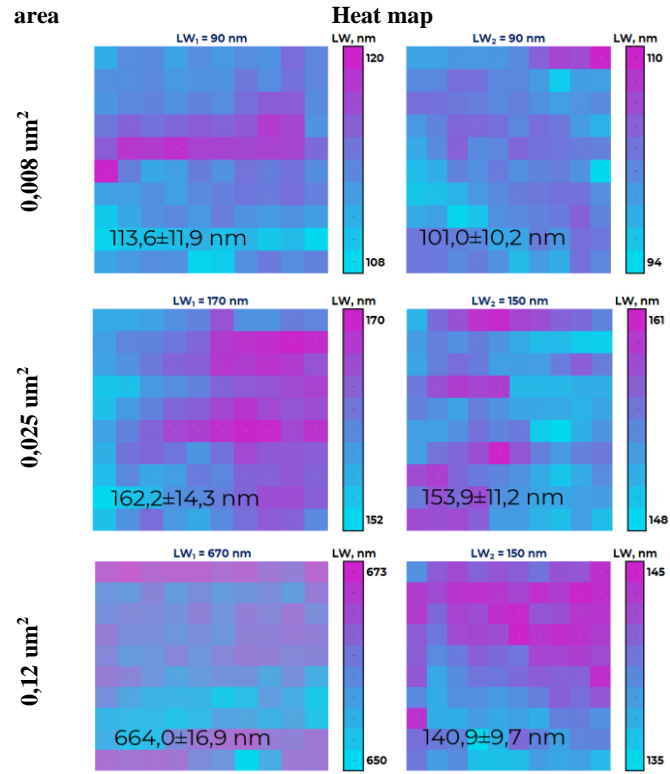